\documentclass[doublecol]{epl2}

\usepackage{graphicx}
\usepackage{latexsym} 
\usepackage[centertags]{amsmath} 
\usepackage{amsfonts} 
\usepackage{amssymb} 
\usepackage{amsthm} 
\usepackage{newlfont} 
\usepackage{subfigure} 
\usepackage{amsbsy} 
\usepackage{color}
\usepackage{bm}

\title{Irreversible dynamics of a massive intruder in dense granular fluids} 

\author{A. Sarracino, D. Villamaina, G. Gradenigo \and A. Puglisi}
\institute{CNR-ISC and Dipartimento di Fisica, Universit\`a Sapienza, p.le A. Moro 2, 00185 Roma, Italy, EU}


\abstract{ A Generalized Langevin Equation with exponential memory is
  proposed for the dynamics of a massive intruder in a dense granular
  fluid. The model reproduces numerical correlation and response
  functions, violating the equilibrium Fluctuation Dissipation
  relations. The source of memory is identified in the coupling of the
  tracer velocity $V$ with a spontaneous local velocity field $U$ in
  the surrounding fluid: fluctuations of this field introduce a
    new timescale with its associated lengthscale. Such
  identification allows us to measure the intruder's fluctuating
  entropy production as a function of $V$ and $U$, obtaining a neat
  verification of the Fluctuation Relation.}

\begin{document} 

\maketitle 

\def\be{\begin{equation}} 
\def\ee{\end{equation}} 

Models of granular fluids are a natural framework where the issues of
non-equilibrium statistical mechanics can be addressed~\cite{JNB96b}. Due to dissipative
interactions among the microscopic constituents, energy is not
conserved and external sources are necessary in order to maintain a
stationary state.  Heat fluxes and currents continuously pass through
the system, time reversal invariance is broken and consequently,
properties such as the Equilibrium Fluctuation-Dissipation relation
(EFDR) do not hold. In recent years, a rather complete theory, at
least in the dilute limit, has been developed and numerous aspects
have been clarified, in good agreement with numerical
simulations~\cite{BP04,BMG09}. However, a general
understanding of dense granular fluids is still lacking. A common
approach is the so-called Enskog correction~\cite{BP04,DS06}, which
reduces the breakdown of Molecular Chaos to a renormalization of
the collision frequency. In cooling regimes, the Enskog
theory may describe strong non-equilibrium effects, due to the explicit
cooling time-dependence~\cite{SD01}. However it cannot describe
dynamical effects in stationary regimes, such as large violations of
the Einstein relation~\cite{G04,PBV07}.

In this letter, we propose a model for the dynamics of a massive
tracer moving in a gas of smaller granular particles, both coupled to
an external bath. In particular, taking as reference point the dilute
limit, where the system has a closed analytical
description~\cite{SVCP10}, we suggest a Generalized Langevin Equation
(GLE) with an exponential memory kernel as first approximation capable
of describing the dense case. Here, the main features are: i) the
decay of correlation and response functions is not simply exponential
and shows backscattering~\cite{OK07,FAZ09} and ii) the
EFDR~\cite{KTH91,BPRV08} of the first and second kind do not hold.  In
the model we propose, detailed balance is not necessarily satisfied,
non-equilibrium effects can be taken into account and the correct
behavior of correlation and response functions is
predicted. Furthermore, the model has a remarkable property: it can be
mapped onto a two-variable Markovian system, i.e. two coupled Langevin
equations with simple white noises. 
The auxiliary variable can be
identified in the local velocity field spontaneously appearing in the
surrounding fluid. This allows us to measure the fluctuating entropy
production~\cite{seifert05}, and fairly verify the Fluctuation
Relation~\cite{Kurchan,LS99,BPRV08}. This is a remarkable
result, if considered the interest of the community~\cite{BGGZ06b} and compared with unsuccessful past
attempts~\cite{FM04,PVBTW05}.

We consider an ``intruder'' disc of mass $m_0=M$ and radius $R$,
moving in a gas of $N$ granular discs with mass $m_i=m$ ($i>0$) and
radius $r$, in a two dimensional box of area $A=L^2$. We denote by
$n=N/A$ the number density of the gas and by $\phi$ the occupied
volume fraction, i.e. $\phi=\pi(Nr^2+R^2)/A$ and we denote by $\bm{V}$
(or $\bm{v}_0$) and $\bm{v}$ (or $\bm{v}_i$ with $i>0$) the velocity
vector of the tracer and of the gas particles, respectively.
Interactions among the particles are hard-core binary instantaneous
inelastic collisions, such that particle $i$, after a collision with
particle $j$, comes out with a velocity
\begin{equation}
\bm{v}_i'=\bm{v}_i-(1+\alpha)\frac{m_j}{m_i+m_j}[(\bm{v}_i-\bm{v}_j)\cdot\hat{\bm{n}}]\hat{\bm{n}}
\end{equation}
where $\hat{\bm{n}}$ is the unit vector joining the particles' centers
of mass and $\alpha \in [0,1]$ is the restitution coefficient
($\alpha=1$ is the elastic case).
The mean free path of the intruder is proportional to $l_0=1/(n(r+R))$ and we denote
by $\tau_c$ its mean collision time. Two kinetic temperatures can be
introduced for the two species: the gas granular temperature
$T_g=m\langle \bm{v}^2\rangle/2$ and the tracer temperature
$T_{tr}=M\langle \bm{V}^2\rangle/2$.

In order to maintain a granular medium in a fluidized state, an
external energy source is coupled to each particle in the form of a
thermal bath~\cite{WM96,NETP99,PLMPV98} (from hereafter,
exploiting isotropy, we consider only one component of the
velocities):
\begin{equation}
m_i\dot{v}_i(t)=-\gamma_b v_i(t) + f_i(t) + \xi_b(t).
\label{langgas}
\end{equation}
Here $f_i(t)$ is the force taking into account the collisions of particle $i$ with other
particles, and $\xi_b(t)$ is a white noise (different for all particles), with
$\langle\xi_b(t)\rangle=0$ and $\langle\xi_{b}(t)\xi_{b}(t')\rangle
=2T_b\gamma_b\delta(t-t')$.  The effect of the external energy source
balances the energy lost in the collisions and a stationary state is
attained with $m_i\langle v_i^2 \rangle \leq T_b$ .

For low packing fractions, $\phi\lesssim 0.1$, and in the large mass
limit, $m/M\ll 1$, using the Enskog approximation it has been shown~\cite{SVCP10} that the dynamics
of the intruder is described by a linear Langevin equation.
In this limit the velocity autocorrelation function shows a simple
exponential decay, with characteristic time $M/\Gamma_E$, where
\begin{equation}
\Gamma_E=\gamma_b+\gamma_g^E, \quad \textrm{with} \quad
\gamma_g^E=\frac{g_2(r+R)}{l_0}\sqrt{2\pi mT_g}(1+\alpha)
\end{equation}
and $g_2(r+R)$ is the pair correlation function for a gas particle and
the intruder at contact. Time-reversal and the EFDR, which are very
weakly modified for uniform dilute granular
gases~\cite{PBL02,G04,PVTW06}, become perfectly satisfied for a
massive intruder. The temperature of the tracer is computed as
$T_{tr}^E=(\gamma_bT_b+\gamma_g^E\frac{1+\alpha}{2}T_g)/\Gamma_E$. For
a general study of a Langevin equation with ``two temperatures'' but a
single time scale (which is always at equilibrium), see
also~\cite{V06}.

As the packing fraction is increased, the Enskog approximation is less
and less effective in predicting the memory effects and the dynamical properties of the
system.  In particular, velocity autocorrelation $C(t)=\langle
V(t)V(0)\rangle/\langle V^2\rangle$ and linear response function
$R(t)=\overline{\delta V(t)}/\delta V(0)$ (i.e. the mean response at time $t$
to an impulsive perturbation applied at time 0)
present an exponential decay modulated in amplitude by oscillating
functions~\cite{FAZ09}.  Moreover violations of the EFDR $C(t)=R(t)$
(Einstein relation) are observed for $\alpha<1$~\cite{PBV07,VPV08}.  

Molecular dynamics simulations of the system have been performed by
means of a standard event driven algorithm to treat hard core
interactions: the algorithm is supplemented with a ``driving event''
at times which are multiples of a small timestep (smaller than all
timescales) which update the velocity of all particles by a
discretized version of Eq.~\eqref{langgas}. In the simulations we have measured $C(t)$ and
$R(t)$, for several different values of the parameters $\alpha$ and
$\phi$. 
\begin{figure}[!htb]
\includegraphics[width=.9\columnwidth,clip=true]{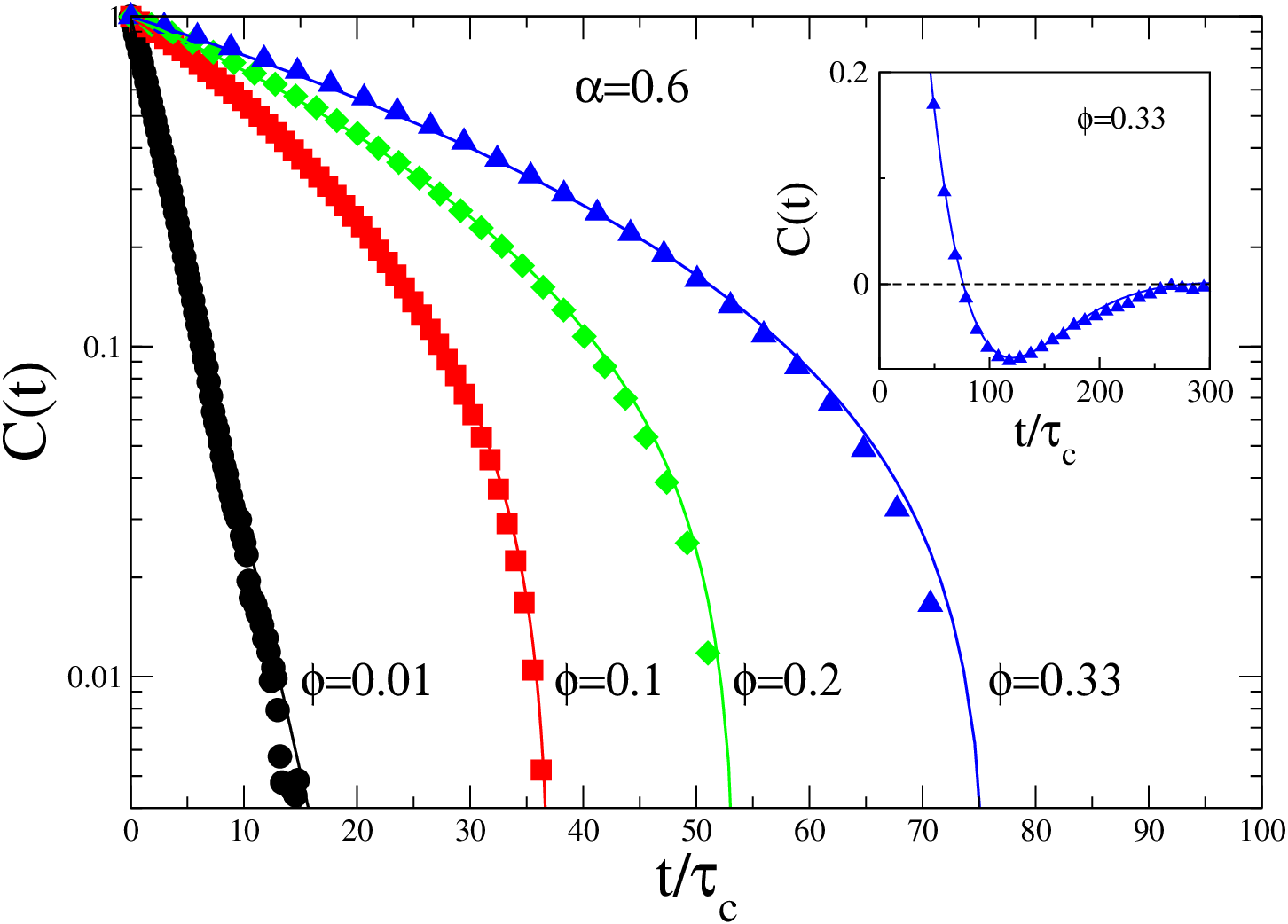}
\caption{(Color online). Semi-log plot of $C(t)$ (symbols) for
  different values of $\phi=0.01,0.1,0.2,0.33$ at $\alpha=0.6$. Times
  are rescaled by the mean collision time $\tau_c$. Continuous lines
  are the best fits obtained with Eq.~(\ref{Cfit}). Inset: $C(t)$ and
  the best fit in linear scale for $\phi=0.33$ and $\alpha=0.6$.}
\label{fig_corr}
\end{figure}
In Fig.~\ref{fig_corr} symbols correspond to the velocity correlation
functions measured in the inelastic case, $\alpha=0.6$, for different
values of the packing fraction $\phi$.  The other parameters are
fixed: $N=2500$, $m=1$, $M=25$, $r=0.005$, $R=0.025$, $T_b=1$,
$\gamma_b=200$. 

Notice that the Enskog
approximation~\cite{BP04,SVCP10} cannot predict the observed functional
forms, because it only modifies by a constant factor the collision
frequency.  In order to describe the full phenomenology, a model with more than one
characteristic time is needed. 
As a first proposal, we consider
a Langevin equation with a single exponential memory kernel~\cite{BBR66,ZBCK05}
\begin{equation} 
M\dot{ V}(t)=-\int_{-\infty}^tdt'~\Gamma(t-t') V(t')+ {\cal E}'(t),
\label{langmemory}
\end{equation}
where 
\begin{equation}
\Gamma(t)=2\gamma_0\delta(t)+\gamma_1/\tau_1e^{-t/\tau_1} 
\end{equation}
and ${\cal E}'(t)={\cal E}_0(t)+{\cal E}_1(t)$, with
\begin{align}
\langle{\cal
  E}_0(t){\cal E}_0(t') \rangle=2 T_0\gamma_0\delta(t-t'),\\
\langle{\cal E}_1(t){\cal E}_1(t') \rangle=T_1\gamma_1/\tau_1e^{-(t-t')/\tau_1}
\end{align}
 and $\langle {\cal E}_1(t){\cal
  E}_0(t') \rangle=0$.  In the limit $\alpha\to 1$, the parameter
$T_1$ is meant to tend to $T_0$ in order to fulfill the EFDR of the
$2$nd kind $\langle {\cal E}'(t){\cal
  E}'(t')\rangle=T_0\Gamma(t-t')$. Within this model the dilute case
is recovered if $\gamma_1\to 0$.  In this limit, the
parameters $\gamma_0$ and $T_0$ coincide with
$\Gamma_E$ and $T_{tr}^E$ of the Enskog theory~\cite{SVCP10}.

\begin{table*}[!htb]
\caption{Parameters of model~(\ref{local_field}), as obtained by fitting
the numerical data (see text for details).  
\label{tabella}}
\begin{center}
\begin{tabular}{|c|c|c|c||c|c|c|c|c||c|c|c|c|}
\hline
$\alpha$ & $\phi$ & $T_{tr}$ & $T_g$ & $\gamma_0/M$ & $T_0$ & $T_1$ & $\gamma_1/M$ & $\tau_1/\tau_c$ & $\Gamma_E/M$ & $\gamma_g^E/M$ & $T_{tr}^E$ & $T_g^E$ \\
\hline
1.0 & 0.33 &  1.00 & 1.00 & 55 & 0.99 & 1.0 & 44 & 67 & 55 & 47 & 1.00 & 1.00 \\
0.8 & 0.33 &  0.92 & 0.90 & 47 & 0.91 & 1.0 & 42 & 68 & 48 & 40 & 0.84 & 0.89 \\
0.7 & 0.33 &  0.88 & 0.86 & 45 & 0.85 & 1.0 & 41 & 74 & 45 & 37 & 0.78 & 0.86 \\
0.6 & 0.33 &  0.86 & 0.84 & 44 & 0.82 & 1.1 & 43 & 89 & 42 & 34 & 0.73 & 0.83 \\
0.6 & 0.20 &  0.92 & 0.91 & 27 & 0.90 & 1.0 & 26 & 54 & 24 & 16 & 0.82 & 0.91 \\
0.6 & 0.10 &  0.95 & 0.96 & 17 & 0.95 & 0.99 & 12 & 29 & 15 & 7 & 0.89 & 0.96 \\
0.6 & 0.01 &  0.99 & 1.00 & 9.6 & 0.99 & /    & 0    & 2.8 & 8.6 & 0.6 & 0.98 & 0.99 \\
0.6 & $0.01^*$ &  0.88 & 0.94 &  21 & 0.88 & /    & 0    & 21 & 20 & 12 & 0.85 & 0.93 \\
\hline
\end{tabular}
\end{center}
\end{table*}

The exponential form of the memory kernel can be justified within the
mode-coupling approximation scheme. In this framework~\cite{HMD96}, it
can be written as a sum of two contributions: $\Gamma(t-t') = \beta_1
\delta(t-t') + \beta_2 \tilde{\Gamma}(t-t')$, where $\beta_1$ and $\beta_2$
are model dependent coefficients, and $\tilde{\Gamma}(t-t')$ is a sum
over modes $q$ of $p(q) e^{-(\nu+D)q^2(t-t')}$, where $p(q)$
weights the modes relevant for the dynamics of the tracer.
Here $D$ and $\nu$ are  the
diffusion coefficient and the kinematic viscosity of the fluid respectively.
Following an old recipe~\cite{BBR66}, tested with success in equilibrium contexts, we assume that, for not too high packing
fractions, memory arises due to re-collisions within a limited region
at distance $\sim \lambda_1$ around the tracer and that this can be
modeled by an effective $p(q)$ which is peaked around
$q_1=2\pi/\lambda_1$, i.e. a single mode contributes to the sum,
yielding $\tilde{\Gamma}(t-t') \sim e^{-(\nu+D) q_1^2 (t-t')}$ and
then
\begin{equation}
\tau_1=\lambda_1^2(2\pi)^{-2}(\nu+D)^{-1}\sim \tau_c^g (\lambda_1/l_0^g)^2,
\label{eq:lambda-tau}
\end{equation}
with $\tau_c^g$ and $l_0^g$ the fluid mean free time and mean free path respectively. Eq.~(\ref{eq:lambda-tau}) relates the time-scale $\tau_1$,
characterizing the tail of the memory kernel, with a typical
length-scale $\lambda_1$ present in the system.  
This length-scale will turn out to play a central role in the
following.

The model~(\ref{langmemory}) predicts $C=f_C(t)$ and $R=f_R(t)$ with
\begin{equation}
f_{C(R)}=e^{-gt}[\cos(\omega t)+a_{C(R)}\sin(\omega t)]. \label{Cfit} 
\end{equation} 
$g$, $\omega$, $a_C$ and $a_R$ are known algebraic functions of
$\gamma_0$, $T_0$, $\gamma_1$, $\tau_1$ and $T_1$. In particular, the
ratio $a_C/a_R=[T_0-\Omega (T_1-T_0)] /[T_0+\Omega (T_1-T_0)]$, with
$\Omega=\gamma_1/[(\gamma_0+\gamma_1)(\gamma_0/M\tau_1-1)]$.  Hence,
in the elastic ($T_1 \to T_0$) as well as in the dilute limit
($\gamma_1 \to 0$), one gets $a_C=a_R$ and recovers the EFDR
$C(t)=R(t)$. In Fig.~\ref{fig_corr} the
continuous lines show the result of the best fits obtained using
Eq.~\eqref{Cfit} for the correlation function, at restitution
coefficient $\alpha=0.6$ and for different values of the packing
fraction $\phi$. The functional form fits very well the numerical
data.

Looking for an insight of the relevant physical mechanisms underlying
such a phenomenology and in order to make clear the meaning of the
parameters, it is useful to map Eq.~(\ref{langmemory}) onto a
Markovian equivalent model by introducing an auxiliary field~\cite{VBPV09}:
\begin{eqnarray}
M\dot{V}&=&-\gamma_0 (V-U)+\sqrt{2T_0\gamma_0}{\cal E}_V \nonumber \\
\dot{U}&=&-\frac{U}{\tau_1}-\frac{\gamma_1}{\gamma_0\tau_1} V
+\sqrt{2\frac{T_1\gamma_1}{\gamma_0^2\tau_1^2}}{\cal E}_U,
\label{local_field}
\end{eqnarray}
where ${\cal E}_V$ and ${\cal E}_U$ are white noises of unitary
variance. The variable
\begin{equation}
U(t)\propto\gamma_1/(\tau_1\gamma_0)\int_{-\infty}^te^{-\frac{t-t'}{\tau_1}}[V(t')+{\cal
  E}_1(t')]dt' 
\label{aux}
\end{equation}
is determined up to a multiplicative factor, as it can be checked by
direct substitution. In the chosen form~(\ref{local_field}), the
dynamics of the tracer is remarkably simple: indeed $V$ follows a
memoryless Langevin equation in a {\em Lagrangian frame} with respect
to a local field $U$. In the dilute limit this is exact (see Appendix
of~\cite{SVCP10}) if $U$ is the \emph{local average velocity field} of
the gas particles colliding with the tracer. Extrapolating such an
identification to higher densities,
we are able to both assign a meaning and predict a
value for most of the parameters of the model: 1) the self drag
coefficient of the intruder in principle is not affected by the change
of reference to the Lagrangian frame, so that $\gamma_0 \sim
\Gamma_E$; 2) for the same reason $T_0 \sim T_{tr}$ is roughly the
temperature of the tracer; 3) $\tau_1$ is the main
relaxation time of the average velocity field $U$ around the Brownian
particle; 4) $\gamma_1$ is the intensity of coupling felt by the
surrounding particles after collisions with the intruder; 5) finally
$T_1$ is the ``temperature'' of the local field $U$, easily
identified with the bath temperature $T_1 \sim T_b$: indeed, thanks to momentum
conservation, inelasticity does not affect the average
velocity of a group of particles almost only colliding among
themselves.

To find a confirmation of the above hypothesis, we have explored the
region of the space of parameters $\alpha\in [0.6,1]$ and
$\phi\in[0.01,0.33]$. From the simultaneous fit of the numerical data
for correlation and response functions against Eqs.~(\ref{Cfit}) we
can determine the set of parameters $\{g,\omega,a_C,a_R,\langle V^2
\rangle\}$. Then, by inverting the relations between them and the set
$\{\gamma_0,T_0,\gamma_1,\tau_1,T_1\}$, we are eventually able to
determine all the parameters entering~(\ref{langmemory}).  
In Table~\ref{tabella} such values are reported, together with the
predictions given by the Enskog approximation (last four columns).
The statistical error on these values is about $1\%$. We used the external
parameters mentioned before, changing $\alpha$ or the box area $A$ (to
change $\phi$): this makes the limit $\phi \to 0$ equivalent to
$\gamma_g \sim 1/l_0 \to 0$ (``super-dilute'' limit). The last row
reports about the true dilute limit: i.e. $R$ is reduced, at fixed
$l_0$ (equal to the value of the previous case $\phi=0.2$), in order
to get $\phi=0.01$ and $\gamma_g > 0$. Notice that in the two dilute
cases the simple Langevin equation is recovered ($\gamma_1=0$) and the
dependence on the parameter $T_1$ disappears.  Remarkably our
predictions $\gamma_0 \sim \Gamma_E$, $T_0 \sim T_{tr}$ and $T_1 \sim
T_b$ are fairly verified. The coupling time $\tau_1$ increases with
the packing fraction and, weakly, with the inelasticity. In the most
dense cases it appears that $\gamma_1 \sim \gamma_g^E \propto \phi$:
this is confirmed in the ``super-dilute'' limit, but cannot hold in
the dilute one, where $\gamma_1 \to 0 \ll \gamma_g^E$. It is also
interesting to notice that at high density $T_{tr} \sim T_g \sim
T_g^E$, which is probably due to the stronger correlations among
particles. Finally we notice that, at large $\phi$, $T_{tr}>T_{tr}^E$,
which is coherent with the idea that correlated collisions dissipate
{\em less} energy.

\begin{figure}[!htb]
\includegraphics[width=.9\columnwidth,clip=true]{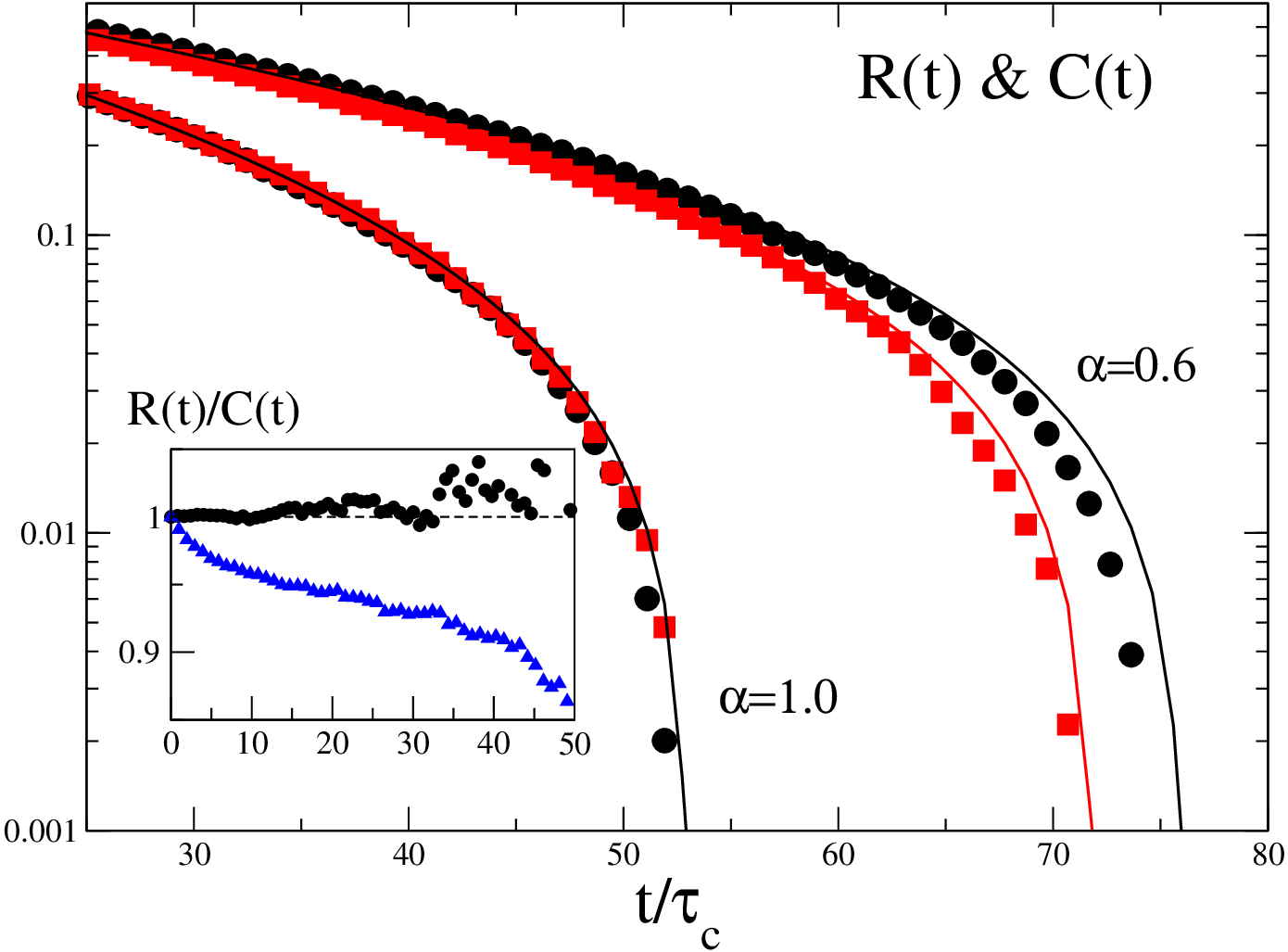}
\caption{(Color online). Correlation function $C(t)$ (black circles) and response function $R(t)$ (red
squares) for $\alpha=1$ and $\alpha=0.6$, at $\phi=0.33$. 
Continuous lines show the best fits curves obtained with Eqs.~(\ref{Cfit}). 
Inset: the ratio $R(t)/C(t)$ is reported in the same cases.}
\label{fig_resp}
\end{figure}

A fundamental feature of this model is its ability to reproduce
violations of EFDR. In Fig.~\ref{fig_resp}, we plot the correlation
and response functions in a dense case (elastic and inelastic):
symbols correspond to numerical data and continuous lines to the best
fit curves.  In the inelastic case, deviations from EFDR $R(t)=C(t)$
are clearly observed. In the inset of Fig.~\ref{fig_resp} the ratio
$R(t)/C(t)$ is also reported.  It is interesting to note that a
relation between the response and correlations measured in the
unperturbed system still exists, but - in the non-equilibrium case -
must take into account the contribution of the cross correlation
$\langle V(t)U(0) \rangle$, i.e.:
\begin{equation}
R(t)=a C(t) + b \langle V(t)U(0)\rangle
\end{equation}
with $a=[1-\gamma_1/M(T_0-T_1)\Omega_a]$ and $b=(T_0-T_1)\Omega_b$,
where $\Omega_a$ and $\Omega_b$ are known functions of the parameters (see for instance~\cite{VBPV09}). At 
equilibrium, where $T_0=T_1$, the EFDR is recorvered.

The mathematical definition of the auxiliary variable $U$,
  Eq.~(\ref{aux}), which requires the knowledge of a part of the noise
  $\cal{E}_1$, makes it very difficult to be measured in simulations
  or in experiments. But the above discussion has shown that $U$
  represents a spontaneous local velocity field interacting with the
  tracer: therefore it can be measured in the following manner. We fix
  a distance $l$ and average the velocity of the gas particles within
  a circle $\cal{C}_l$ of radius $l+R$ centered on the tracer. In this
  way we define $U_l=1/N_l\sum_{i\in\cal{C}_l}v_i$, where $N_l$ is the
  number of particles in $\cal{C}_l$. Two methods are available to
  estimate the correct length $l^*$, which is difficult to be
  predicted on a general ground. A first guess is provided by
  identifying it with $\lambda_1$, which can be obtained by inverting
  Eq.~(\ref{eq:lambda-tau}) after having measured $\tau_1$, using the
  known values of $D$ and $\nu$ in a granular fluid. The second method
  is to measure the correlations $\langle VU_l\rangle$ and $\langle
  U_l^2\rangle$ and find the best value $l_{cor}$ such that $\langle
  VU_{l_{cor}}\rangle\sim \langle VU\rangle$ and $\langle
  U_{l_{cor}}^2\rangle\sim\langle U^2\rangle$ (where $\langle
  VU\rangle$ and $\langle U^2 \rangle$ are easily computed from the
  model, once all the parameters have been determined fitting $C(t)$
  and $R(t)$). Remarkably, the two estimates give compatible results
  and identify a narrow range of values for $l^*\sim \lambda_1 \sim
  l_{cor}$. Hence, one can identify $U\sim U_{l^*}$ and the auxiliary
variable can be directly measured in numerical simulations and experiments.

An important independent assessment of the effectiveness of
model~\eqref{langmemory} comes from the study of the fluctuating
entropy production~\cite{seifert05} which quantifies the deviation
from detailed balance in a trajectory. Given the trajectory in the
time interval $[0,t]$, $\{V(s)\}_0^t$, and its time-reversed $\{{\cal
I}V(s)\}_0^t\equiv\{-V(t-s)\}_0^t$, in Ref.~\cite{PV09} it has been
shown that the entropy production for the model~(\ref{langmemory})
takes the form
\begin{equation}
\Sigma_t=\log\frac{P(\{V(s)\}_0^t)}{P(\{{\cal I}V(s)\}_0^t)}
\approx \gamma_0\left(\frac{1}{T_0}-\frac{1}{T_1}\right)\int_0^t ds~V(s)U(s).
\label{entropy_prod}
\end{equation}
Boundary terms - in the stationary state - are subleading for large
$t$ and have been neglected.  This functional vanishes exactly in the
elastic case, $\alpha=1$, where equipartition holds, $T_1=T_0$, and is
zero on average in the dilute limit, where $\langle VU\rangle=0$.
Formula~\eqref{entropy_prod} reveals that the leading source of
entropy production is the energy transferred by the ``force''
$\gamma_0 U$ on the tracer, weighed by the difference between
the inverse temperatures of the two ``thermostats''.
\begin{figure}[!htb]
\includegraphics[width=.9\columnwidth,clip=true]{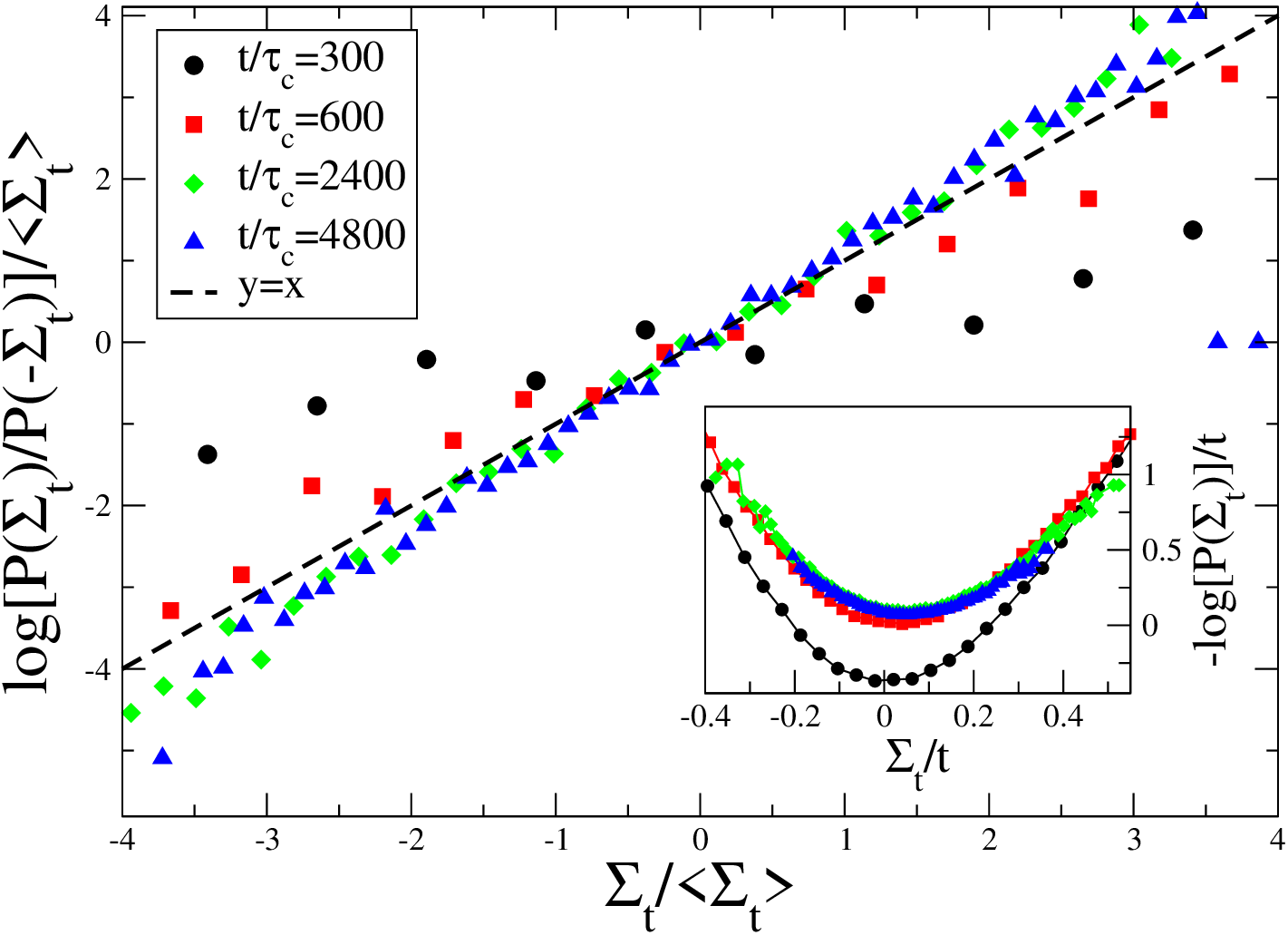}
\caption{(Color online). Check of the fluctuation
relation~(\ref{GCrel}) in the system with $\alpha=0.6$ and
$\phi=0.33$. 
Inset: collapse of the rescaled probability
distributions of $\Sigma_t$ at large times onto the large deviation function.}
\label{GC}
\end{figure}
Following the procedure described above, in the case
$\phi=0.33$ and $\alpha=0.6$, we estimate for the
correlation length $l^*\sim 9r \sim 6l_0$. Then, measuring 
 the entropy production of Eq.~(\ref{entropy_prod}) (by replacing $U(t)$ with $U_{l^*}$)
along many trajectories of length $t$, we can
 compute the probability $P(\Sigma_t=x)$ and compare it to
$P(\Sigma_t=-x)$, in order to verify the Fluctuation Relation
\begin{equation}
\log\frac{P(\Sigma_t=x)}{P(\Sigma_t=-x)}=x.
\label{GCrel}
\end{equation}
In Fig.~\ref{GC} we report our numerical results. The main frame
confirms that at large times the Fluctuation Relation~(\ref{GCrel}) is
well verified within the statistical errors.  The inset shows the
collapse of $\log P(\Sigma_t)/t$ onto the large deviation rate
function for large times.  Notice also that formula~\eqref{entropy_prod}
does not contain further parameters but the ones already determined by
correlation and response measure, i.e. the slope of the graph is not
adjusted by further fits. Indeed a wrong evaluation of the weighing
factor $(1/T_0-1/T_1) \approx(1/T_{tr}-1/T_b)$ or of the ``energy
injection rate'' $\gamma_0 U(t) V(t)$ in Eq.~\eqref{entropy_prod}
could produce a completely different slope in Fig.~\ref{GC}.

In conclusion, we designed a first granular dynamical theory
describing non-equilibrium correlators and responses for a massive
tracer. The value of this proposal is to offer a significant insight
into the mechanisms of re-collision and dynamical memory and their
unexplored relation with the breakdown of equilibrium properties. It
is remarkable that velocity correlations $\langle V(t) U(t') \rangle$
between the intruder and the surrounding velocity field are
responsible for both the violations of the EFDR and the appearance of
a non-zero entropy production, provided that the two fields are {\em
at different temperatures}.  Small non-Gaussian
corrections~\cite{PVTW06}, always present in granular fluids, are
neglected here in favor of the largest contribution given by memory
terms to violations of EFDR and entropy production. For some of the
parameters in the theory ($\gamma_0 \sim \Gamma_E$, $T_0 \sim T_{tr}$
and $T_1 \sim T_b$) we have reasonable predictions, while $\tau_1$ and
$\gamma_1$, related to the coupling between $U$ and $V$, deserve
further investigations. Close analytical predictions of all the
parameters could be obtained through a full kinetic theory (beyond
Enskog), also to deduce eventual extensions to the case $M \sim m$,
larger densities, and hard spheres.

\begin{acknowledgments} We thank A.~Vulpiani and P.~Visco 
  for a careful reading of the manuscript. The work is
  supported by the ``Granular-Chaos'' project, funded by the Italian
  MIUR under the FIRB-IDEAS grant number RBID08Z9JE.
\end{acknowledgments}

\bibliographystyle{eplbib}

\bibliography{fluct.bib}

\end{document}